  \documentclass[multi]{cambridge6Atight}
  \usepackage{natbib}

  \usepackage{rotating}
  \usepackage{floatpag}
  \rotfloatpagestyle{empty}

  \usepackage{amsthm}
  \usepackage{graphicx}
  \usepackage{mathptmx}
  
  \usepackage{longtable}

  \usepackage{makeidx}
  \makeindex

\copyrightline{This material has been published in \textit{The Impact of Binaries on Stellar Evolution}, Beccari G. \& Boffin H.M.J. (Eds.).
This version is free to view and download for personal use only. Not for re-distribution, re-sale or use in derivative works. \copyright\ 2018 Cambridge University Press.}
    
\begin{document}

%%%%%%%%%%%%%%%%%%%%%%%%%%
%%%% START TO EDIT FROM HERE %%%%%%
%%%%%%%%%%%%%%%%%%%%%%%%%%

  \alphafootnotes
   \author[David Jones]
    {David Jones}
  %%%% PUT HERE TITLE OF YOUR CHAPTER  
  \chapter{The importance of binarity in the formation and evolution of planetary nebulae}
  %%% Put A. Einstein et al. in [] if more than 3 authors - these are running authors
  %%% \chapter[Einstein et al.]{Writing a chapter for the Imbase17 book}

%%% footnotes are not compulsory... please use only if needed. Don't put your affiliation here.
%  \footnotetext[1]{Supported by my wife, under grant 12345}
  \arabicfootnotes

%%% Provide here again for all authors of the chapter, the name and affiliation -- only one author at the time!
  \contributor{David Jones
    \affiliation{Instituto de Astrof\'isica de Canarias,
    E-38205 La Laguna, Tenerife, Spain}}

%%%% PUT HERE YOUR ABSTRACT
 \begin{abstract}
It is now clear that a binary evolutionary pathway is responsible for a significant fraction of all planetary nebulae, with some authors even going so far as to claim that binarity may be a near requirement for the formation of an observable nebula.  In this chapter, we will discuss the theoretical and observational support for the importance of binarity in the formation of planetary nebulae, initially focussing on common envelope evolution but also covering wider binaries.  Furthermore, we will highlight the impact that these results have on our understanding of other astrophysical phenomena, including supernovae type Ia, chemically peculiar stars and circumbinary exoplanets.  Finally, we will present the latest results with regards to the relationship between post-common-envelope central stars and the abundance discrepancy problem in planetary nebulae, and what further clues this may hold in forwarding our understanding of the common envelope phase itself.
 \end{abstract}
 
\section{Introduction}
Planetary nebulae (PNe)\index{Planetary nebula} are glowing shells of gas and dust surrounding (pre-)white dwarfs\index{White dwarf} (pre-WD).  The canonical formation scenario for PNe is a single star of low to intermediate initial mass, which has recently evolved off the AGB (and is thus a pre-white dwarf), ionising its own slow-moving AGB ejecta that has been swept up into a shell by a fast, tenuous wind originating from the central star \citep{kwok78}.  This model, frequently referred to as the Interacting Stellar Winds (ISW) model\index{Interacting stellar winds model}, successfully reproduces a wide-range of PN properties, however it fails to account for the plethora of elaborate, often highly axisymmetric, morphologies observed \citep{balick02}.  

Many additional ingredients to the ISW model have been considered, including rapid rotation on the AGB and strong magnetic fields in single stars \citep{garcia-segura14,nordhaus07}, however only one remains a viable candidate for producing the most axisymmetric structures found in PNe -- central star binarity \citep[e.g.][]{jones17b}.  The idea that some PNe might host binary central stars is not a new one, with such systems being considered key evidence for the existence of the common envelope\index{Common envelope} (CE) phase first proposed in 1976 by \citet{paczynski76} and the discovery of the first post-CE binary central star shortly after \citep{bond76}.  However, it is only recently that the importance of binarity, not only in the shaping of PNe but also in their formation, has begun to be truly appreciated.

In a wider context, many binary phenomena experienced by low- and intermediate-mass stars represent evolutionary phases that follow the production of a PN, meaning that PNe with binary central stars offer an important window to study the origins of such objects including Barium stars, cataclysmic variables, stellar mergers, novae and, perhaps most importantly, the cosmologically important supernovae type Ia.
 
\section{Close-binary systems}
\label{djones:closebinpn}
While many forms of binary interaction will influence the evolution of the system, particularly in terms of the stellar mass-loss rates, none will have such a dramatic effect as the occurrence of a CE phase.  Similarly, the implications of a CE phase on the large scale morphology of any resulting PN are perhaps the most easily understood.  Via the conservation of angular momentum, the ejected CE will always be preferentially deposited into the binary orbital plane resulting in a clearly axisymmetric structure.  It then stands to reason that the interaction between the ejected envelope and a subsequent wind, originating from the now exposed pre-WD core, will form a similarly axisymmetric structure, providing a natural explanation for bipolar PNe \citep{nordhaus06}.  

\subsection{Binary fraction}
\label{djones:binfracpn}
Perhaps the clearest obstacle to accepting the importance of central star binarity as the key ingredient in the shaping of PNe has been the lack of a robust measure of the binary fraction amongst the central stars of PNe (CSPNe).  Several estimates of the close-binary fraction had been made by compiling the results of multiple surveys, each of which employing different observing techniques with differing sensitivities making accounting for biases impossible \citep[e.g.][]{bond00}, but it was not until the OGLE survey that a more definitive fraction could be derived.  \citet{miszalski09a} used photometry from the OGLE microlensing survey to search for photometric variability in the central stars of some 149 PNe, deriving a detectable close-binary fraction of 12--21\% and more than doubling the sample of known binary CSPN in the process.  This figure can be considered a hard lower limit to the true binary fraction given that it only represents the binary fraction detectable by the survey \citep{jones17b}, which is highly sensitive to secondary type (lower mass secondaries will go undetected), orbital separation/period (long period, wider binaries will go undetected) as well as apparent magnitude (fainter CSPNe or those contaminated by bright nebulae were not recoverable).  However, this fraction is indeed capable of accounting for the most axisymmetric PNe, with a similar fraction of the total PN population showing such bipolar morphologies \citep{frew16}.
  
  \begin{center}
  \begin{longtable}{p{2cm}p{3cm}p{1.5cm}p{4.5cm}}
 \caption[Post-CE PNe]
      {List of post-CE PNe with confirmed orbital periods (a more detailed version is continually maintained by the author at http://drdjones.net/bCSPN ).}
    \label{djones:postcepntab}\\

       \hline \hline
	PN G & Common name & Period (days) & Reference\\
        \hline
\endfirsthead

	\multicolumn{4}{c}{\tablename\ \thetable{} -- continued from previous page}\\
       \hline
	PN G & Common name & Period (days) & Reference\\
        \hline
        \endhead
        
        \hline \multicolumn{4}{r}{Continued on next page} \\ \hline
        \endfoot
        
        \hline \hline
        \endlastfoot
  000.2-01.9	&	M 2-19 			&0.67	&	\cite{miszalski09a}	\\
000.5-03.1a	&	MPA J1759-3007 	&0.50	&	\cite{miszalski09a}	\\
000.6-01.3	&	Bl 3-15 			&0.27	&	\cite{miszalski09a}	\\
000.9-03.3	&	PHR J1801-2947 	&0.32	&	\cite{miszalski09a}	\\
001.2-02.6	&	PHR J1759-2915 	&1.10	&	\cite{miszalski09a}	\\
001.8-02.0	&	PHR J1757-2824 	&0.80	&	\cite{miszalski09a}	\\
001.9-02.5	&	PPA J1759-2834 	&0.31	&	\cite{miszalski09a}	\\
005.0+03.0	&	Pe 1-9 			&0.14	&	\cite{miszalski09a}	\\
005.1-08.9	&	Hf 2-2 			&0.40	&	\cite{hillwig16a}	\\
009.6+10.5	&	Abell 41 			&0.23	&	\cite{jones10}	\\
017.3-21.9	&	Abell 65 			&1.00	&	\cite{hillwig15}	\\
034.5-06.7	&	NGC 6778		&0.15	&	\cite{miszalski11b}	\\
049.4+02.4	&	Hen 2-428 		&0.18	&	\cite{santander-garcia15}	\\
053.8-03.0	&	Abell 63 			&0.46	&	\cite{afsar08}	\\
054.2-03.4	&	The Necklace		&1.16	&	\cite{corradi11}		\\
055.4+16.0	&	Abell 46			&0.47	&	\cite{afsar08}\\
058.6-03.6	&	Nova Vul 2007		&0.07	&	\cite{rodriguez-gil10}	\\
068.1+11.0	&	ETHOS 1			&0.53	&	\cite{miszalski11a} \\
075.9+11.6	&	AMU 1			&2.93	&	\cite{demarco15}	\\
076.3+14.1	&	Patchick 5			&1.12	&	\cite{demarco15}	\\
086.9-03.4	&	Ou 5 			&0.36	&	\cite{corradi14}	\\
135.9+55.9	&	TS01 			&0.16	&	\cite{tovmassian10}	\\
136.3+05.5	&	HFG 1			&0.58	&	\cite{exter05}	\\
144.8+65.8	&	LTNF 1 			&2.29	&	\cite{ferguson99}	\\
-			&	GK Per			&1.90	&	\cite{bode87}	\\
215.6+03.6	&	NGC 2346		&15.99	&	\cite{brown19}\\
220.3-53.9	&	NGC 1360		&142		&	\cite{miszalski17}\\
222.8-04.2	&	PM 1-23			&1.26	&	\cite{manick15}\\
242.6-11.6        &	M 3-1  			&0.13        &	\cite{jones19}\\
253.5+10.7	&	K 1-2			&0.68	&	\cite{exter03}\\
259.1+00.9	&	Hen 2-11			&0.61	&	\cite{jones14}\\
283.9+09.7	&	DS 1 			&0.36	&	\cite{hilditch96}\\
290.5+07.9	&	Fg 1 				&1.20	&	\cite{boffin12}\\
307.2-03.4	&	NGC 5189 		&4.04	&	\cite{manick15}\\
307.5-04.9	&	MyCn 18			&18.15	&	\cite{miszalski18}\\
%316.7-05.8	&	MPA J1508-6455 	&12.50	&	\\
329.0+01.9	&	Sp 1				&2.91	&	\cite{hillwig16b}\\
%331.5-02.7	&	Hen 2-161 		&$\sim$1	&	\cite{jones15}\\
332.5-16.9	&	HaTr 7 			&0.32	&	\cite{hillwig16c}\\
335.2-03.6	&	HaTr 4 			&1.74	&	\cite{hillwig16a}\\
337.0+08.4	&	ESO 330-9 		&0.30	&	\cite{hillwig16c}\\
338.1-08.3	&	NGC 6326 		&0.37	&	\cite{miszalski11b}\\
338.8+05.6	&	Hen 2-155 		&0.15	&	\cite{jones15}\\
341.6+13.7	&	NGC 6026 		&0.53	&	\cite{hillwig10}\\
%349.3-04.2	&	Lo 16 			&0.49	&	\\
349.3-01.1	&	NGC 6337 		&0.17	&	\cite{hillwig10}\\
354.5-03.9	&	Sab 41 			&0.30	&	\cite{miszalski09a}\\
355.3-03.2	&	PPA J1747-3435 	&0.23	&	\cite{miszalski09a}\\
355.7-03.0	&	Hen 2-283 		&1.13	&	\cite{miszalski09a}\\
357.0-04.4	&	PHR J1756-3342 	&0.27	&	\cite{miszalski09a}\\
357.1-05.3	&	BMP J1800-3407 	&0.15	&	\cite{miszalski09a}\\
357.6-03.3	&	H 2-29 			&0.24	&	\cite{miszalski09a}\\
358.7-03.0	&	K 6-34 			&0.20	&	\cite{miszalski09a}\\
358.8+04.0	&	Th 3-15 			&0.15	&	\cite{soszynski15}\\
359.1-02.3	&	M 3-16 			&0.57	&	\cite{miszalski09a}\\
359.5-01.2	&	JaSt 66 			&0.28	&	\cite{miszalski09a}\\

\end{longtable}
\end{center}

Surveys based on, for example, radial velocity variability \citep[more sensitive to longer orbital periods, e.g.][]{demarco04} or infrared excesses due to the contribution of a lower temperature companion to the total spectral energy distribution of the CSPN \citep[more sensitive to lower mass companions, e.g.][]{douchin15} tend to derive much higher binary fractions (approaching 100\%) but with far greater uncertainties.  The Kepler mission, which provides space-based photometry with a precision impossible for a ground-based survey, offers a unique opportunity to probe further the photometrically detectable binary fraction.  The first Kepler field contained only five CSPNe with useable data, however four of these were found to present variability consistent with binarity (or, rather, binary evolution -- one was determined to be a merger product), indicative that an analysis of data from later fields containing more CSPNe may lead to a similarly large binary fraction \citep{demarco15}.  Consequently, it is now clear that the binary fraction amongst CSPNe is appreciable, at least $\sim$20\% but probably even greater.  Furthermore, this fraction does not include systems that were initially binary but merged during the CE, the total number of which is rather uncertain, however there is some evidence that such systems do go on to form PNe with morphologies similar to those that survived the CE \citep[][and references therein]{miszalski12b}.  Ultimately, the total number of known post-CE binary CSPNe with solid period determinations is a little over fifty, with all those systems being listed in table \ref{djones:postcepntab} (with a more detailed version continually maintained by the author at http://drdjones.net/bCSPN ).

\subsection{The link between morphology and central star binarity}
\label{djones:morphpn}

\begin{figure}
    \includegraphics[height=7cm]{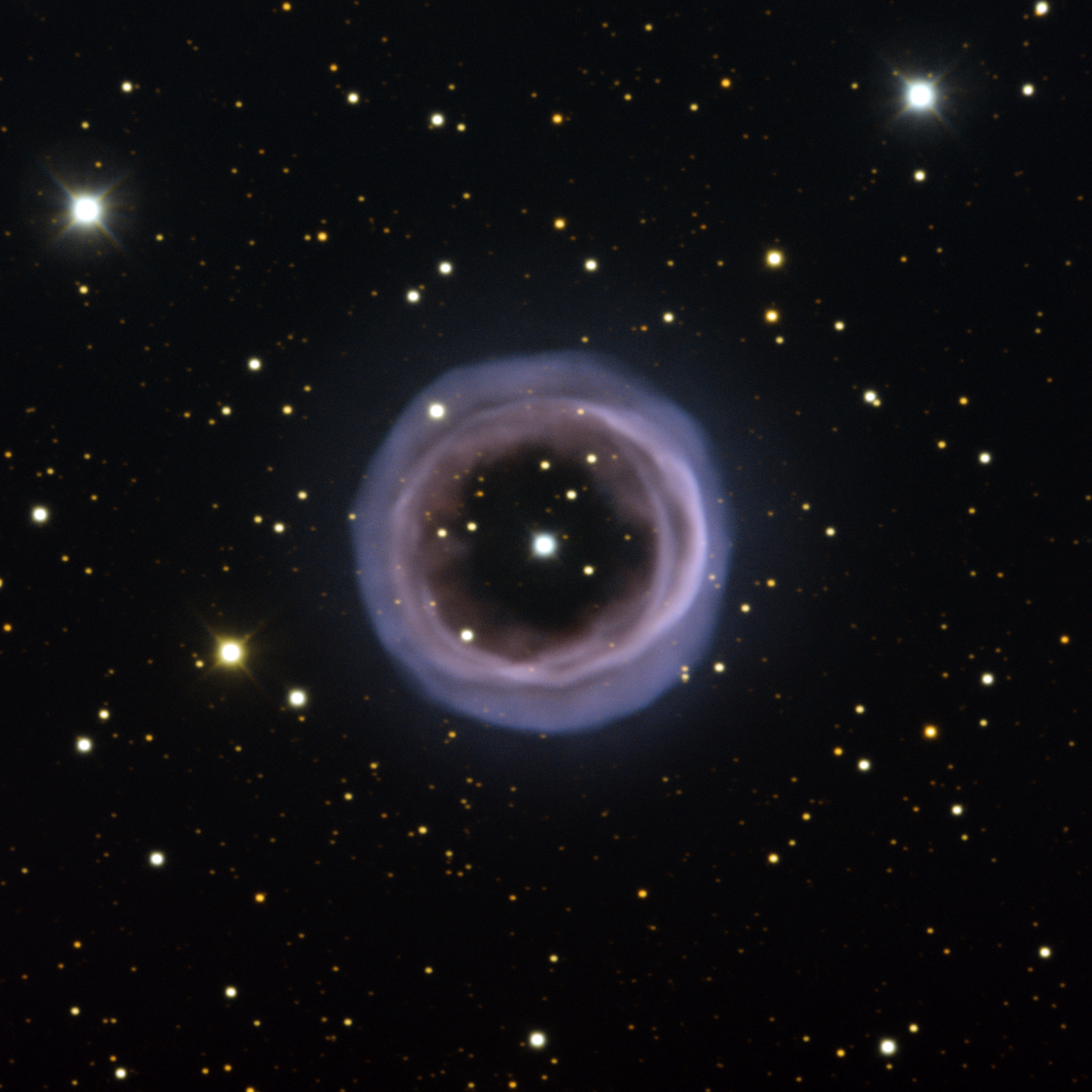}
    %  note that the square brace option below is only required
    %  if you intend to produce a list of illustrations
    \caption[Image of PN Shapley 1]
      {A colour composite image of PN Shapley 1 (image credit: ESO).  Upon first inspection, the PN appears to present a roughly spherical structure but is, in fact, a near-pole-on bipolar, the inclination of which is consistent with that of its central binary star \citep{jones12,hillwig16b}.}
    \label{djones:sp1}
     \end{figure}

With the greatly increased number of known binary CSPNe following the work of \citet{miszalski09a}, it was possible to assess the sample for prevalent morphological characteristics which might reasonably be used to set them apart from the rest of the PN population (i.e.\ those where no evidence of binarity was detected).  Just as expected from the models of CE ejection, post-CE PNe show a strong tendency to present with canonical bipolar structures \citep[likely more than half are bipolar once inclination effects are accounted for;][]{miszalski09b}.  This strong connection between bipolarity and central star binarity has since been confirmed beyond doubt - in \emph{every} case, where the two have been derived, the nebular symmetry axis is found to lie perpendicular to the orbital plane of the binary \citep{hillwig16b}, including in the Fine Ring Nebula (Sp~1; figure \ref{djones:sp1}) which was initially believed to be spherical but has since been shown to be a near pole-on bipolar structure \citep{jones12}.

\begin{figure}
    \includegraphics[height=5cm]{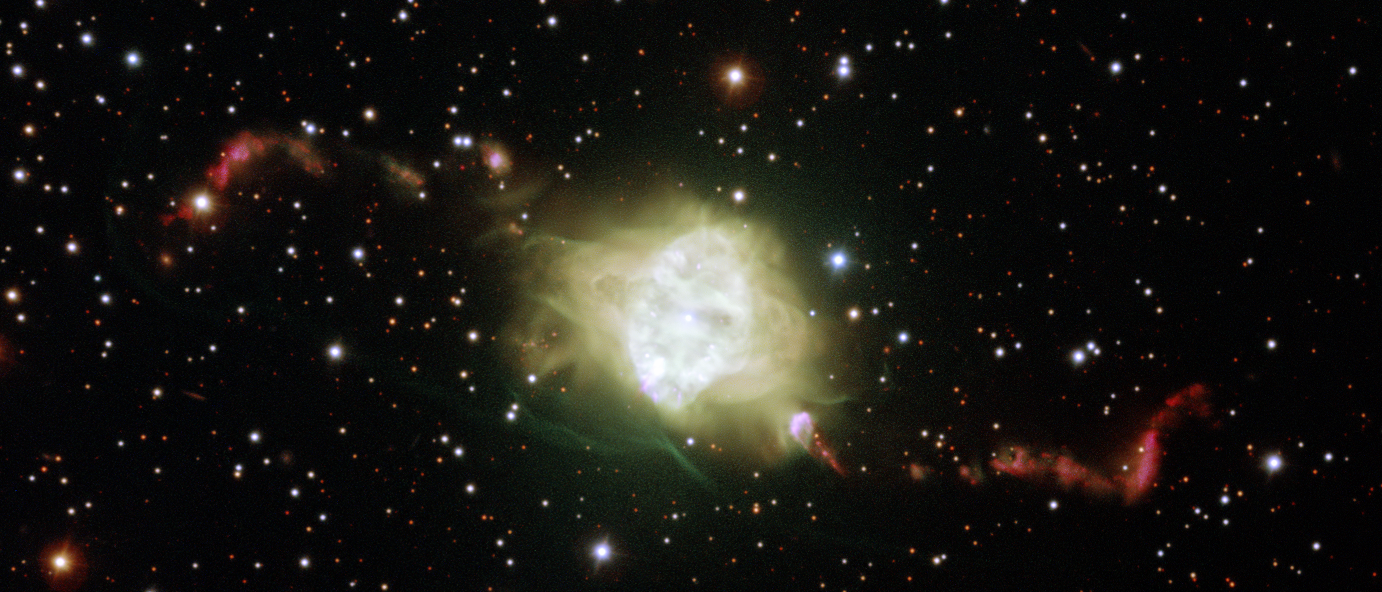}
    %  note that the square brace option below is only required
    %  if you intend to produce a list of illustrations
    \caption[Image of PN Fleming 1]
      {A colour composite image of PN Fleming 1 showing its remarkable pair of precessing jets, which are thought to have been produced by an intense episode of mass transfer just prior to the CSPN entering the CE phase \citep{boffin12}. Image credit: ESO/H.~Boffin.}
    \label{djones:fleming1}
     \end{figure}

\begin{figure}
    \includegraphics[height=5cm]{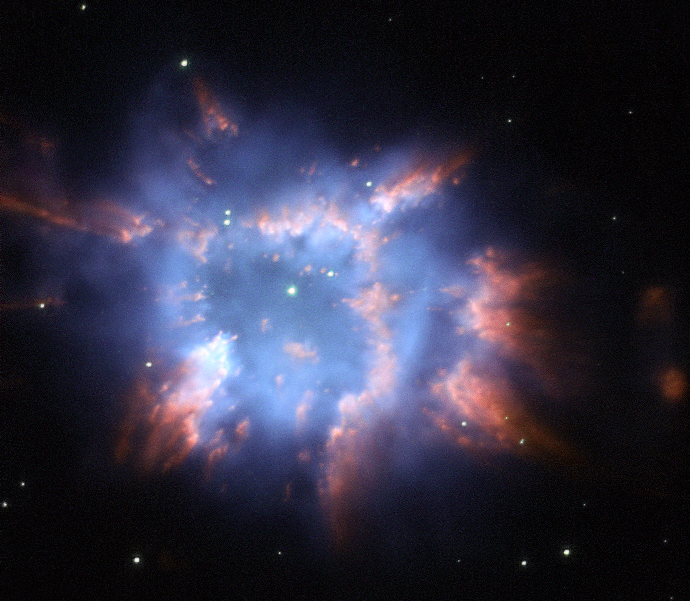}
    %  note that the square brace option below is only required
    %  if you intend to produce a list of illustrations
    \caption[Image of PN NGC~6326]
      {A colour composite image of PN NGC~6326 highlighting the filamentary structures found by \citet{miszalski09b} to be typical of central star binarity \citep{miszalski11b}. Image credit: ESA/Hubble and NASA.}
    \label{djones:ngc6326}
     \end{figure}

Intriguingly, \citet{miszalski09b} also found a strong correlation between central star binarity and the presence of equatorial rings/torii, jets\index{Jet} and low-ionisation filamentary structures (see figures \ref{djones:fleming1} and \ref{djones:ngc6326}).  Equatorial torii \citep[e.g.\ Abell 41,][]{jones10} and knotty waists \citep[e.g.\ The Necklace,][]{corradi11} can clearly be explained under the same prescription as the bipolar structures, if the amount of CE material deposited in the orbital plane is sufficiently high \citep[knots can arise from the interaction of a photo-ionising wind with an inhomogeneous, clumpy torus;][]{miszalski09b}.  Similarly, jets are a natural result of mass transfer between the binary components, and while most hydrodynamical models predict that accretion onto the secondary is minimal during the CE \citep{ricker12,macleod15} there is now growing evidence that a significant amount of material might be accreted before entering into the CE (see section \ref{precepn}).  Low-ionisation filaments not aligned in either the polar or equatorial directions, such as those of NGC~6326 \citep[see figure \ref{djones:ngc6326} and][]{miszalski11b}, are perhaps more difficult to understand, but are thought to form through a wind-shell interaction whereby modest inhomogeneities in the shell (which would, in these systems, be the ejected CE) are enhanced by the influence of an encroaching fast, low-density wind \citep{steffen13}. 

In order to yet further increase the known binary CSPN population, whilst minimising the required investment of observing time, these newly identified morphological features have all been used to select PNe for targeted follow-up leading to an accelerated rate of discovery \citep[e.g.][]{miszalski11a,miszalski11b, jones14, jones15,jones19}.  As the sample of known binary stars has grown (as well as the sample that have been the subject of detailed studies based on multi-band photometric and radial velocity observations), several interesting surprises have been revealed, some of which will be discussed in sections \ref{djones:sniapn} and \ref{precepn}.

%%%% EXAMPLE OF A FIGURE %%%%%

\subsection{The link to supernovae type Ia}
\label{djones:sniapn}

Perhaps the most intriguing result that can be derived from the current sample of known post-CE CSPNE is the apparent over-abundance of binary systems comprised of two white dwarfs, with approximately 20\% of all known post-CE CSPNe found to be such double degenerate\index{Double degenerate} (DD) binaries \citep{jones17b}.  This high fraction of post-CE DD CSPNe is at odds with population synthesis models, perhaps indicating that the expected birthrate of such systems could be much higher than previously suspected \citep{demarco08}.  Such a finding has important consequences for our understanding of the CE phase given that many of these systems would have had two survive CEs \citep{tovmassian10}.  Under the classical prescription of the CE, whereby some fraction ($\alpha$) of the orbital energy is used to unbind and eject the envelope of the giant star \citep{demarco11}, it stands to reason that surviving two such events presents more of a challenge.  For low mass companions, this prescription shows that orbital energy alone would not be enough to eject the envelope during even the first CE, indicating that some other energy source, thought by some to be the recombination energy of the envelope itself, is required \citep{nandez16}.  For more massive companions, the ejection of the first CE is easily understood, but how the resulting spiralled-in binary could survive a second CE is more challenging to explain, but cannot be excluded given that the energy released for a given change in separation is much greater at shorter periods.  It is, however, problematic that, due to the now shortened orbital period, the second CE is significantly more likely to occur when the companion is ascending the RGB (rather than the AGB), when its envelope is more bound.  It has been suggested that, under certain circumstances, the first CE can be avoided with the initially more massive component of the binary losing the majority of its envelope via a period of stable, non-conservative mass transfer, and the system then evolving to DD through a single CE \citep{woods12}.  Similarly, a grazing envelope (GE) evolution, whereby the giant radius and orbital separation shrink simultaneously, could also provide a means to unbind the envelope of either (or both) star(s) allowing the system to become DD before merging \citep{soker15}.

Studying DD CSPNe offers a unique window into the formation of these systems as the presence of a PN indicates that they have only recently left the final (or perhaps only) CE phase of their evolution, and that the system has not yet had the time to adjust thermally.  Furthermore, understanding the formation processes for DD binaries has great implications in the study of supernovae type Ia\index{Supernova Ia} (SNe Ia) given that the merger of such systems is one of the proposed pathways for their formation \citep{iben84}.  Indeed, several recent studies have suggested that a DD merger may be responsible for most, if not all, SNe Ia \citep{maoz14,maoz17}, an hypothesis which is perhaps supported most strongly by the observed delay-time distribution of observed SNe Ia \citep{maoz10}.  Understanding the origins of these SNe Ia is of critical importance given their role in the discovery of the accelerated expansion of the Universe \citep{schmidt98, reiss98, perlmutter99}, which was ultimately awarded the 2011 Nobel Prize for Physics.  However, in spite of their cosmological importance, it is still unclear why SNe Ia present with the same brightness  \citep[once corrected for the shape of their post-explosion light curves;][]{phillips93}, and hence how they can be such astonishingly precise standard candles.  

The best candidate SN Ia progenitor found to date is the CSPN of Hen~2-428, shown by \citet{santander-garcia15} to comprise a DD binary, the total mass of which is above the Chandrasekhar limit and which will merge in less than a Hubble time.  Other DD CSPNe are also found to have short merging times but with a total mass less than the Chandrasekhar mass \citep[e.g. TS 01;][]{tovmassian10}.  These may not result in classical SNe Ia, but will clearly be observed as transients and perhaps produce stellar classes such as R Coronae Borealis\index{R Coronae Borealis star} stars \citep{clayton11}.  Beyond the DD scenario for SN Ia formation, it is important to note that some of the strongest candidates for the single degenerate (SD) pathway, whereby a massive WD reaches the Chandrasekhar mass via accretion from a companion, are also found inside PNe \citep[e.g. Nova Vul 2007;][]{rodriguez-gil10}.   Finally, a significant fraction of SNe Ia are found to explode in circumstellar environments consistent with the presence of a previously ejected PN \citep{tsebrenko15}, indicating that, irrespective of the dominant evolutionary pathway towards SNe Ia, CSPNe will be key to understanding their formation (as well as those of other astrophysical transients and merger products).

\subsection{Pre-common-envelope mass transfer}
\label{precepn}

Further to the apparent over-abundance of DD systems, studies of the single degenerate systems - those comprised of a WD and a main sequence (MS) star - have also brought surprising results.  All systems with a MS secondary, that have been the subject of detailed modelling based on both light and radial velocity curves, are found to present with companions that are inflated with respect to their field star counterparts \citep{jones15}.  While the total number of systems modelled is rather low, this is still a striking result given the levels of inflation (in some cases, the measured radii are a factor of two or more greater than those predicted by zero age main sequence models, see table \ref{djones:pnradii}). It is now generally accepted that this inflation is a result of rapid mass transfer either during or just prior to the CE phase.  This rapid mass transfer knocks the secondaries out of thermal equilibrium causing their radii to grow, and given that their thermal timescales (millions of years) are much larger than the timescales for CE ejection and PN visibility (tens of thousands of years) they remain out of equilibrium.  The high levels of irradiation in these systems, as highlighted by the tendency of these MS stars to present with higher temperatures than single star counterparts (see table \ref{djones:pnradii}), probably also act to maintain the secondaries out of thermal equilibrium but are unlikely to be the primary cause of the inflation itself \citep{demarco08}.

  \begin{table}[h!]
    \begin{minipage}{180pt}
    %  note that the square brace option below is  required
    %  as we intend to produce a list of tables
    \caption[Parameters of main sequence secondaries in post-CE CSPNe]
      {Parameters of main sequence secondary stars in post-common-envelope planetary nebula central stars \citep[][and references therein]{jones15}.}
    \label{djones:pnradii}
    %%%% Replace Author by your last name please %%%%
%%%% example can be replaced by what you want %%%%
    \addtolength\tabcolsep{2pt}% to stretch columns, if required
      \begin{tabular}{@{}l@{\hspace{25pt}}ccc@{}}
        \hline \hline
	PN & Mass & Radius & Temperature\\
	      & (M$_\odot$) & (R$_\odot$) & (kK)\\
        \hline
	Abell 46 	& 0.15$\pm$0.02 & 0.46$\pm$0.02 & 3.9$\pm$0.4\\
	Abell 63 	& 0.29$\pm$0.03 & 0.56$\pm$0.02 & 6.1$\pm$0.2\\
	Abell 65 	& 0.22$\pm$0.04 & 0.41$\pm$0.05 & 5.0$\pm$1.0\\
	DS 1 	& 0.23$\pm$0.01 & 0.40$\pm$0.01 & 3.4$\pm$1.0\\
	Hen 2-155& 0.13$\pm$0.02 & 0.30$\pm$0.03 & 3.5$\pm$0.5\\
	LTNF 1 	& 0.36$\pm$0.07 & 0.72$\pm$0.05 & 5.8$\pm$0.3\\	
        \hline \hline
      \end{tabular}
    \end{minipage}
  \end{table}

Further support for this hypothesised phase of intense mass transfer comes from the observed jet-like structures in many post-CE PNe.  Jets are believed to be a natural consequence of accretion and, as such, the clear connection between these structures and central star binarity can be considered a strong indication that significant accretion has occurred \citep{miszalski09b,tocknell14,sowicka17}.  The secondary in the Necklace nebula (which shows prominent jet-like structures) has been shown to be a carbon dwarf\index{Carbon dwarf}, presenting significant contamination of AGB processed material which must have been accreted from the evolved primary \citep[][further discussion of chemical contamination in binary CSPNe can be found in Section \ref{djones:bapn}]{miszalski13a}.  Additionally, the kinematical ages of the jets (including those of the Necklace) also tend to be older than those measured for the central regions of these PNe suggesting that the jets are launched prior to entering the CE \citep{corradi11}. This chronology is supported by observations of Fleming 1, where the precession period of the jets indicates that the orbital period of the binary at the time of ejection was much larger than the current post-CE period \citep[i.e.\ before the CE in-spiral;][]{boffin12}.  In a few cases, there are suggestions that the jets are formed after the ejection of the CE (or, in the case of some proto-PNe, contemporaneously), perhaps indicative of multiple mass transfer episodes or even fallback \citep{tocknell14,jones16,jones17b}.  This idea will be revisited in Section \ref{djones:abundpn}.

Many binary CSPNe with MS companions have been observed to be X-ray point sources \citep{kastner12,freeman14}, with most displaying relatively hard spectra that are unlikely to arise due to photospheric emission from the host CSPNe themselves \citep{montez15}. Instead, the emission is thought to originate from coronal emission from the spun-up main-sequence companion \citep{montez10}.  Given the evolved nature of these systems, this activity is almost certainly a consequence of binary interaction \citep[magnetic activity in late-type main sequence stars decreases with age;][]{soker02}, but it is unclear whether the spin-up is due to accretion or merely tidal locking in these short period systems \citep{montez15}.  

\section{Non-post-common-envelope systems}
\label{nonpostce}

Section \ref{djones:closebinpn} focussed on post-CE PNe, however even binaries that avoid the CE might be expected to have a significant impact on the formation of a PN.  Indeed, some authors have even gone so far as to claim that the entire PN population may derive from binary stars, once wider binaries are included, and that most single stars may be precluded from forming an observable PN \citep[e.g.][]{moe06}.  This idea is somewhat supported by the high binary fractions derived from methodologies sensitive to such wider binaries (see Section \ref{djones:binfracpn}), however more recent models of post-AGB evolution strongly indicate that single stars should be capable of producing observable PNe \citep{mmmb16}.

In the following subsections, we will discuss some special cases amongst the PNe known to host wide-binary central stars, and their implications for the importance of such systems in the PN population as a whole.

\subsection{Barium stars}
\label{djones:bapn}

Section \ref{precepn} discussed mass transfer as part of the CE, but several binary configurations will experience stable mass transfer and avoid the orbital shrinkage associated with a CE.  One such class of objects are the Barium stars\index{Barium star}, consisting of a WD and a cool (spectral type G to K), giant companion that has been chemically enriched with Barium and other s-process elements \citep{mcclure80}.  Given that the companions are not evolved enough to have produced such significant quantities of s-process elements, the chemical enrichment is believed to be the consequence of accreting s-process enriched material from the primary, most likely via wind Roche-lobe overflow\index{Wind Roche-lobe overflow} \citep{boffin88,theuns96}.  The wind Roche-lobe overflow mechanism - whereby, while the primary never fills its Roche lobe its stellar wind does - allows for greatly increased accretion rates compared to the classical Bondi-Hoyle-Lyttleton accretion prescription.  If the acceleration radius of the primary's wind is comparable or greater than its Roche-lobe radius, the wind will be strongly influenced by the binary potential and be accreted onto the secondary via the first Langrangian point.

Several Barium star CSPNe have been discovered \citep[e.g.][]{miszalski12a, miszalski13b} though, to date, only one has a measured orbital period \citep[LoTr~5 at a somewhat typical period for this class of object, $\sim$7.5 years;][]{aller18}.  In some cases, the rotation rate of the giant secondary has been measured via photometric variability indicating that they are highly spun-up by the mass transfer \citep[rotation periods of the order of a few days;][]{tyndall13}.  Furthermore, analyses of the PNe surrounding these Barium stars shows a tendency towards toroidal structures \citep{tyndall13}, indicative that mass loss through the system is also enhanced through the second and third Langrangian points, as predicted by smoothed particle hydrodynamics simulations of wind Roche-lobe overflow \citep{mohamed12}.  In this regard, Ba CSPNe offer a unique window into this poorly understood mass transfer process, given that the observed PN represents the remnant of the wind which escaped the binary potential.

\subsection{Long-period radial velocity variables}

A small number of long-period binary CSPNe have been discovered through long-term radial velocity monitoring campaigns.  These sorts of campaigns have only been made possible by the new generation of high stability, high resolution spectrographs which permit the long-term monitoring of stellar sources with a precision of a few metres per second.  Given the need for continued access to such facilities over long time periods, surveys for such long-period CSPNe have generally been limited to smaller telescopes greatly restricting the number of sufficiently bright candidates.  However, in spite of these limitations, one survey with the HERMES spectrograph mounted on the 1.2m Mercator Telescope has revealed three long-period binary CSPNe with periods in the range 3--10 years \citep{vanwinckel14,jones17a}.  In all three cases, the brightest component of the binary is the secondary star \citep[in most cases a giant, though their is some doubt over the nature of the secondary of NGC~1514;][]{jones17a} providing some indication that these systems represent only the ``tip of the iceberg'' when it comes to such long-period CSPNe, and that similar studies with larger telescopes (thus probing fainter binaries) should unearth many more such systems.  Furthermore, the CSPN of the longest period of these systems, that of NGC~1514, was previously the subject of a long-term monitoring campaign lasting almost one year but which failed to reveal its binary nature due to the its extreme eccentricity \citep[the one year campaign was unfortunate enough to observe the CSPN at phases when the variability was at a minimum;][]{jones17a}.  In light of this, previous monitoring campaigns which have found few variables cannot be considered conclusive as many long-period systems would evade detection (either due to the velocity precision of the instrumentation used, or the duration of the monitoring programmes).  This, compounded with other surveys which have indicated a large number of radial velocity variable CSPNe but which failed to find periodicities \citep[e.g.][]{demarco04}, is strongly indicative of a large number of long-period binary CSPNe which have, to date, evaded detection.

\begin{figure}
    \includegraphics[height=5cm]{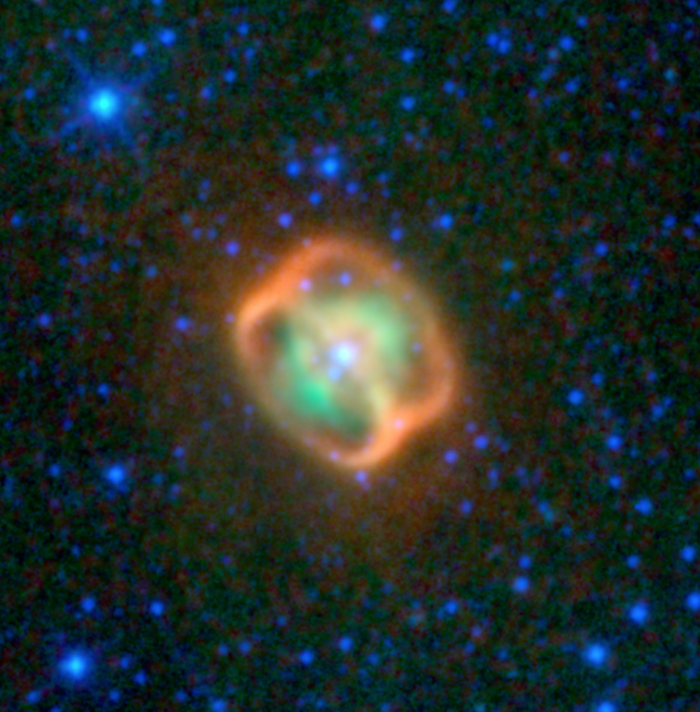}
    %  note that the square brace option below is only required
    %  if you intend to produce a list of illustrations
    \caption[Image of PN NGC~1514]
      {A colour composite image of the infrared emission from PN NGC~1514 highlighting the prominent rings, believed to represent the ends of a bipolar, hourglass-like structure - typical of the influence of a central binary star. Image credit:  NASA/JPL-Caltech/Wise team.}
    \label{djones:ngc1514}
     \end{figure}

Given that there clearly should be an appreciable number of long-period binary CSPNe, the question becomes: does the evolution of these binaries vary significantly from isolated field stars and what does that mean for the formation and/or evolution of any PNe resulting from theses systems?  In the orbital period range of the three known systems, it is unlikely that any has experienced direct Roche-lobe overflow (i.e.\ one component filling its Roche lobe).  However, they are clearly in the regime where wind Roche-lobe overflow could have an appreciable effect.  The observed morphologies of the host PNe support the idea that their binary nature has significantly influenced the mass-loss evolution of the systems, with two of the three showing clearly axisymmetric, bipolar morphologies \citep[NGC~1514 and LoTr~5;][and references therein]{jones17a}, while the morphology of the third (PN~G052.7+50.7) is unclear due to its rather small extent \citep[5'',][]{napiwotzki94}.  NGC~1514 in particular, which hosts the longest period CSPN at $\sim$9 years, presents with a remarkable pair of dust rings (see figure \ref{djones:ngc1514}) somewhat similar to those observed in nebulae surrounding other binary phenomena such as symbiotic stars \citep[e.g.\ Hen~2-104, The Southern Crab;][]{santander-garcia04} and even the type II SN 1987A \citep{morris07}.  Additionally, two of the three long-period CSPNe show stellar chemical abnormalities clearly associated with a binary evolution - the CSPN of LoTr~5 is a Barium star (see section \ref{djones:bapn}) while the CSPN of PN~G052.7+50.7 shows a depletion in refractory elements similar to that found in many post-AGB binaries with circumbinary discs \citep[][and references therein]{napiwotzki94,vanwinckel14}.

\subsection{Resolved companions}

Beyond those few wide binaries for which orbital periods have been derived directly, there are also several CSPNe for which distant companions have been resolved.  \citet{ciardullo99} observed 113 PNe as part of a Hubble Space Telescope ``snapshot'' survey, finding 10 CSPNe with stars so close to their nuclei that their associations are very likely while a further six have possible associations (i.e.\ a resolved binary fraction of $\sim$10\%).  The measured separations of these candidates equate to orbital periods roughly in the range of several hundreds to hundreds of thousands of years, a regime in which most are likely to have experienced little to no interaction even during the nebular progenitors' AGB ascent.  In some cases, this hypothesis is supported by the observed morphologies of the host nebulae, with several presenting (near-)spherical morphologies (e.g.\ Abell 33, K1-14) and observable haloes as predicted by single star models \citep[e.g.\ NGC1535;][]{corradi03}.  The majority, however, present aspherical morphologies, some clearly as a result of interaction with the interstellar medium \citep[e.g.\ Abell 31;][]{wareing07} but most are better explained by a binary interaction.   It is perhaps possible that those with the shortest periods may be close enough to have experienced somewhat enhanced mass-loss rates via the so-called companion-reinforced attrition process \citep[CRAP;][]{tout88} or perhaps even some wind Roche-lobe overflow (see Section \ref{djones:bapn}), but most would be considered too wide.  This raises the intriguing possibility that these systems are (or were), in fact, triple stars\index{Triple star system} wherein the PN progenitor forms a closer binary with an unresolved companion while the resolved companion is the third component of a hierarchical triple \citep{bear17}.  Interestingly, in the only known triple CSPN, all three components are resolvable from the ground using adaptive optics imagery and, as such, the progenitor is unlikely to have experienced any strong binary interaction \citep{adam14}.  To date, no system where the nebular progenitor forms part of a close-binary inside a hierarchical triple has been clearly detected \citep{jones17c} - perhaps the best remaining candidate is Abell 63 which is known to host a post-CE binary central star with a possible third companion at around 3500~AU separation, however high contrast imaging and astrometry are required to confirm the association \citep{ciardullo99,adam14}.

\section{Chemistry}
\label{djones:chempn}

\subsection{Dual-dust chemistry}
\label{djones:dualdustpn}

The dust chemistry of PNe is dominated by the interplay between carbon and oxygen, with the less abundant species locked away in the form of carbon monoxide leaving the more abundant to drive the nebular dust chemistry. Therefore, the majority of PNe are found to display either carbon- or oxygen-rich chemistry in the form of polycyclic aromatic hydrocarbons (PAHs, in the case of carbon-rich dust) or silicates (in the case of oxygen-rich dust).  However, a significant fraction of PNe are found to display so-called dual-dust chemistry\index{Dual-dust chemistry} wherein both PAH and silicate emission features are observed in their spectra.  It had been suggested that these objects may be transition objects, in a phase where the star is transitioning from O-rich to C-rich \citep[following a thermal pulse, for example;][]{guzman-ramirez15}, however the discovery of several dual-dust PNe in the Galactic bulge (where the population of old, low mass stars is not expected to have experienced significant third dredge-up and therefore should not present with the necessary enhanced C/O ratios) throws this hypothesis into doubt \citep{perea-calderon09,garcia-rojas17}.  \citet{guzman-ramirez11} demonstrated that PAHs could be formed in oxygen-rich environments via the photodissociation of carbon monoxide in a UV-irradiated, dense torus, such as those associated with post-CE binary CSPNe (see section \ref{djones:morphpn}).  Mid-infrared imaging of dual-dust PNe supports this hypothesis with PAH emission being restricted to the denser equatorial regions \citep{guzman-ramirez14}.

The strong association between post-CE nebular morphologies (equatorial rings) and dual-dust chemistry provides a clear indication that complex molecular material can be formed almost immediately following the CE phase.  This is particularly interesting in the context of the discovery of circumbinary exoplanets\index{Exoplanet (circumbinary)} around post-CE systems \citep[e.g.\ NN Ser;][]{marsh14}, demonstrating that it is not necessary for these exoplanets to have survived the CE but rather that they may have formed from the remnant of the ejected envelope \citep{schleicher15}.  Bearing this in mind, the name PN may not be such a misnomer given that the nebular material may go on to form second-generation planets around post-CE cores.

\subsection{The abundance discrepancy problem}
\label{djones:abundpn}

\begin{figure}
    \includegraphics[height=5.2cm]{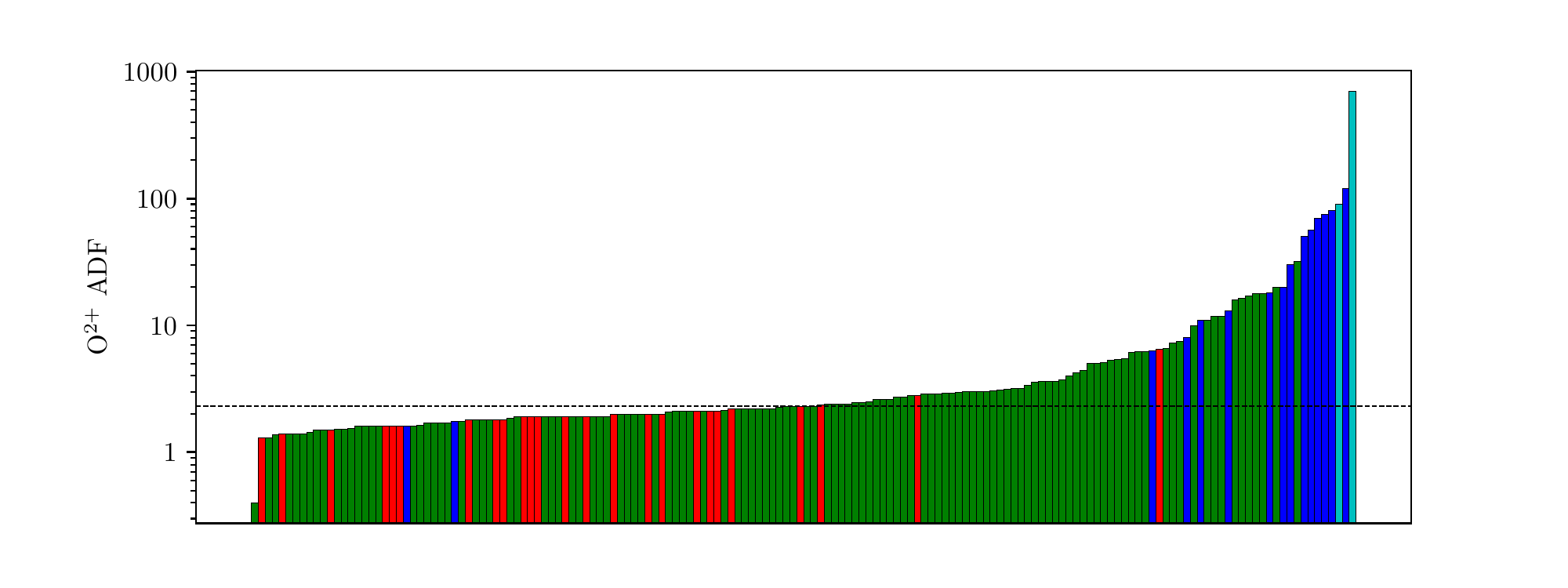}
    %  note that the square brace option below is only required
    %  if you intend to produce a list of illustrations
    \caption[A compilation of O$^{2+}$ ADFs]
      {A compilation of O$^{2+}$ ADFs from the literature including H\textsc{ii} regions (red), PNe (green) PNe with close binary CSPNe (blue) and PNe with born-again CSPNe (cyan).  The horizontal dashed line marks the median value of all measured ADFs at 2.3, this value becomes 1.9 when considering only H\textsc{ii} regions and rises to 2.6 for PNe alone.  Based on data compiled by Roger Wesson ({https://www.nebulousresearch.org/adfs/}).
.}
    \label{djones:adfs}
     \end{figure}

The spectra of PNe present a plethora of emission lines, both as a result of recombination and collisional excitation, from a wide range of elements.  The intensities of these emission lines vary as a function of both the abundance of that particular chemical species and the physical conditions (electron density and temperature) of the emitting gas.  However, for more than seventy years there has been a known discrepancy between the chemical abundances calculated using lines of the same species but with different emission mechanisms \citep{wyse42}. This discrepancy has generally been characterised by an abundance discrepancy factor\index{Abundance discrepancy} (ADF) defined as
\begin{equation}
\mathrm{ADF}(X^{+i}) = \frac{X^{+i}_{\mathrm{RLs}}}{X^{+i}_{\mathrm{CELs}}}
\end{equation}
where $\mathrm{ADF}(X^{+i}) $ is the ADF of a given ion $X^{+i}$, and $X^{+i}_{\mathrm{RLs}}$ and $X^{+i}_{\mathrm{CELs}}$ are the ionic abundances of that species derived using recombination lines (RLs) and collisionally excited lines (CELs), respectively.  The value of the ADF is almost universally, across all forms of astrophysical nebulae and irrelevant of the chosen ion, determined to be greater than unity \citep[see figure \ref{djones:adfs} and ][]{mcnabb13}, representing one of the greatest unresolved problems in nebular astrophysics today \citep{peimbert17}.  Recently, the issue has been further complicated by the discovery that the ADFs measured in PNe with post-CE CSPNe far exceed those of the general population of ionised nebulae \citep[although there are a few exceptions, e.g.\ IC~4776;][]{sowicka17}, with values reaching up to 800 in the central regions of some objects \citep{liu06,corradi15,wesson18}. 

Several explanations have been put forward to explain the ADF problem, including various forms of inhomogeneity (temperature, chemical and density), non-Maxwellian electron velocity distributions and time-dependent photoionisation variations, with the true cause probably comprising some combination of all these factors \citep[see, for example, the review of ][]{peimbert17}.  While the root cause across all ionised nebulae cannot have a binary origin (H\textsc{ii} regions do not have an associated binary), it is clear that the impact of, at least, one of the proposed mechanisms driving the discrepancy is greatly amplified by a close-binary evolution in the CSPN.  Spatially-resolved ADF studies have indicated that in post-CE PNe there does appear to be a strong chemical inhomogeneity, with centrally peaked ADFs being associated with the presence of a second, higher metallicity (and, therefore, lower temperature) gas phase at the core of these nebulae \citep{garcia-rojas16,jones16}.  Alongside the post-CE PNe, the highest ADF values are found in born-again PNe - those systems where it is believed that the CSPN underwent a very late thermal pulse\index{Very late thermal pulse}, ejecting chemically-enriched material into the old nebula \citep{wesson08}.  This evolutionary scenario offers a clear explanation for the presence of a second gas phase in the inner-regions of these nebulae, and perhaps offers insight into the origins of the high ADFs in post-CE PNe given that these born-again PNe are also strongly suspected to host binary CSPNe \citep{wesson08}.  Indeed, it has been suggested that chemical reprocessing of CE material which has fallen-back onto the CSPN could prompt a form of born-again evolution in the post-CE PNe offering a formation mechanism for the chemically-enriched, second gas phase in these nebulae \citep{jones16}.  Further exploration of this hypothesis is essential to understand the evolution of these systems through and beyond the CE, with spatio-kinematical observations in both ORLs and CELs combined with hydrodynamic simulations essential to evaluate the possible formation scenarios of these post-CE PNe and the origin of their high ADFs.

\section{Summary}

Central star binarity has long been thought to play a role in the shaping of PNe, but it is only now that the magnitude of their importance is becoming clear.  Some 20\% of all PNe host post-CE CSPNe that are detectable from the ground via photometric monitoring, with those nebulae showing marked morphological signatures of their binary evolution (bipolar structures, jets, and equatorial rings).  In \emph{all} cases, where both are known, the nebular symmetry axes are found to lie perpendicular to the orbital plane of the binary, just as predicted by models of PN formation by binary interaction - the probability of finding such a correlation by chance is less than one in a million.  Surveys employing different methodologies (radial velocity monitoring, infrared excesses, space-based photometry) indicate that this 20\% binary fraction may well be a dramatic under-estimate, with some determinations (that will include contributions from non-post-CE CSPNe) consistent with a near-100\% fraction.  It is also clear that binary systems that avoid a CE evolution can still have a dramatic impact on the resulting PNe with those hosting long-period (years to tens of years) CSPNe also showing marked departures from sphericity, likewise those few PNe suspected to host the products of CE mergers. Even PNe hosting resolved binaries, with separations so wide that interaction should have been minimal, seem to show morphologies that are difficult to explain in a single-star scenario.  There are still many open questions but it is now clear that central star binarity plays a crucial role in the evolution of a significant fraction of, and perhaps even all, PNe.

PNe with binary central stars offer an important window into a multitude of astrophysical processes, including the common-envelope phase, wind Roche-lobe overflow, launching of jets around stellar mass objects, and the chemical contamination of stellar atmospheres, to name just a few.  They also represent the progenitors of a wide-range of astronomical phenomena including chemically peculiar stars (such as Barium stars and Carbon stars), cataclysmic variables, novae, SNe Ia (irrelevant of their formation mechanism), R Coronae Borealis stars, low-mass X-ray binaries, and, most likely, many other astronomical transients.  Studies of PNe clearly have much to inform on the formation and evolution of these phenomena.  PNe may yet have the final word in the debate over the formation of circumbinary exoplanets around post-CE systems with first generation planets having to survive the PN phase, and second generation planets perhaps being formed from material associated with the PN itself.

Perhaps the most obvious role that PNe can play in wider context is in forwarding our understanding of the CE phase.  Studies have already revealed several aspects to the phase that will be critical in moving forward.  The presence of jets, as well as inflated and chemically-polluted main-sequence secondaries, strongly points to a significant amount of material accreted onto the companion, most-likely just prior to entering the CE phase.  The broad-scale morphologies of post-CE PNe themselves offer a unique opportunity to study the structure of the ejected envelope as the nebula itself is formed from this material.  Post-CE CSPNe represent the immediate products of the CE phase, with the PN ensuring that the binary systems are ``fresh out of the oven'', and, as such, their mass and period distributions are clearly those which CE population synthesis studies should be required to match.  The abundance patterns in post-CE PNe, in particular the high ADFs and the distribution and kinematics of the CEL and RL emitting gas, have much to inform on the CE process, perhaps indicating that the final stages of the process result in the ejection of chemically-enriched material into the already dissipating envelope.  Understanding the CE and its products represent a critical issue in modern astrophysics, and a crucial aspect in our understanding of all binary evolution and particularly in the production of detectable gravitational wave sources as well as cosmologically important SNe Ia.

%\newpage

  \bibliographystyle{cambridgeauthordate}
  
 %%%%%%%%%%%%%%%%%%%%%%%%%%% 
 \copyrightline{} 
 \printindex
    %this is to check if you are happy with your index -- it will not appear here in the book and you can ignore the additional page
    
%%%%%%%%%% END %%%%%%
\end{document}